\documentclass[aoas,preprint]{imsart}

\usepackage[OT1]{fontenc}

\usepackage{microtype}

\usepackage{amsmath,amssymb,amsthm}

\usepackage{bm,bbm,mathtools}

\usepackage{float,graphics,graphicx,subfigure}

\usepackage{url}

\usepackage{natbib}
\let\cite=\citep

\usepackage{array,booktabs,multirow,rotating}

\let\oldmin\min
\let\oldmax\max

\DeclareMathOperator{\I}{I}
\DeclareMathOperator{\bigO}{\mathcal{O}}
\DeclareMathOperator{\arctantwo}{arctan2}

\newcommand{\N}{\mathbb{N}}

\newcommand{\R}{\mathbb{R}}

\newcommand{\abs}[1]{\left\lvert #1 \right\rvert}

\newcommand{\norm}[1]{\left\lVert #1 \right\rVert}
\newcommand{\round}[1]{\left\lfloor #1 \right\rceil}
\newcommand{\floor}[1]{\left\lfloor #1 \right\rfloor}

\renewcommand{\min}[1][]{\underset{#1}{\oldmin}\;}
\renewcommand{\max}[1][]{\underset{#1}{\oldmax}\;}
\newcommand{\argmin}[1][]{\underset{#1}{\text{arg} \oldmin}\;}
\newcommand{\argmax}[1][]{\underset{#1}{\text{arg} \oldmax}\;}

\newcommand{\set}[1]{\left\lbrace #1 \right\rbrace}
\newcommand{\bracket}[2][]{\left\lbrack #2 \right\rbrack_{#1}}
\newcommand{\RG}{\mathbf{R}}                       
\newcommand{\PC}{\mathbf{P}}                       
\newcommand{\CH}{\PC_\text{CH}}                    
\newcommand{\BB}{\PC_\text{MBB}}                   
\newcommand{\CC}{\mathbf{c}}                       
\newcommand{\K}{\mathbf{\Delta c}}                 
\newcommand{\RL}{\mathbf{r}}                       
\newcommand{\NR}{\mathbf{\tilde{r}}}               
\newcommand{\SR}{\mathbf{\hat{r}}}                 

\newcommand{\iRG}[1][wl]{R_{#1}}                    
\newcommand{\iPC}[1][i]{P_{#1\cdot}}                
\newcommand{\iCH}[1][i]{P_{\text{CH}_{#1\cdot}}}    
\newcommand{\iBB}[1][i]{P_{\text{MBB}_{#1\cdot}}}   
\newcommand{\iCC}[1][i]{c_{#1}}                     
\newcommand{\iK}[1][i]{{\Delta c}_{#1}}             
\newcommand{\iRL}[1][i]{r_{#1}}                     
\newcommand{\iNR}[1][i]{\tilde{r}_{#1}}             
\newcommand{\iSR}[1][i]{\hat{r}_{#1}}               

\newcommand{\cent}{P_c}                            
\newcommand{\HR}{\mathbf{H}}                       
\begin{document}
    \begin{frontmatter}
        %
        \title{Discovering \\ 
               Clinically Meaningful Shape Features for the Analysis of \\ 
               Tumor Pathology Images}
        \runtitle{Shape Analysis of Tumor Pathology Images}
        %
        
        \thankstext{T1}{Address correspondence to {qiwei.li@utdallas.edu}.}
        
        \begin{aug}
            %
            \author{\fnms{Esteban} 
                    \snm{Fern\a'andez Morales}%
                    \thanksref{m1}\ead[label=e1]{}},
            \author{\fnms{Cong}
                    \snm{Zhang}%
                    \thanksref{m1}\ead[label=e2]{}},
            \author{\fnms{Guanghua}
                    \snm{Xiao}%
                    \thanksref{m2}\ead[label=e3]{}},
            \author{\fnms{Chul}
                    \snm{Moon}%
                    \thanksref{m3}\ead[label=e4]{}}, \and
            \author{\fnms{Qiwei}
                    \snm{Li}%
                    \thanksref{m1,T1}\ead[label=e5]{}}
            %
            
            
            \runauthor{Fern\a'andez Morales et al.}
            
            \affiliation{The University of Texas at Dallas\thanksmark{m1}, 
                         The University of Texas Southwestern Medical Center\thanksmark{m2}, and 
                         Southern Methodist University\thanksmark{m3}}
            
            %
            \address{Esteban Fern\a'andez Morales, Cong Zhang, and Qiwei Li \\
            	     800 W Campbell Rd \\
            	     Mathematical Sciences, FO 35 \\
            	     The University of Texas at Dallas \\
            	     Richardson, TX 75080, United States \\
            	     \printead{e1} \\
                     \printead{e2} \\
                     \printead{e5}}
            \address{Guanghua Xiao \\
                     5323 Harry Hines Blvd \\
                     Quantitative Biology Research Center, Suite H9.124 \\
                     The University of Texas Southwestern Medical Center \\
                     Dallas, TX 75390, United States \\
                     \printead{e3}}
            \address{Chul Moon \\
            	     3225 Daniel Ave \\
            	     104 Heroy Science Hall \\
            	     Southern Methodist University \\
            	     Dallas, TX 75275, United States \\
            	     \printead{e4}}
            %
        \end{aug}

        \begin{abstract}
            With the advanced imaging technology, digital pathology imaging of tumor tissue slides is becoming a routine clinical procedure for cancer diagnosis. This process produces massive imaging data that capture histological details in high resolution. Recent developments in deep-learning methods have enabled us to automatically detect and characterize the tumor regions in pathology images at large scale. From each identified tumor region, we extracted $30$ well-defined descriptors that quantify its shape, geometry, and topology. We demonstrated how those descriptor features were associated with patient survival outcome in lung adenocarcinoma patients from the National Lung Screening Trial ($n=143$). Besides, a descriptor-based prognostic model was developed and validated in an independent patient cohort from The Cancer Genome Atlas Program program ($n=318$). This study proposes new insights into the relationship between tumor shape, geometrical, and topological features and patient prognosis. We provide software in the form of \texttt{R} code on GitHub: https://github.com/estfernandez/Slide\_Image\_Segmentation\_and\_Extraction.
        \end{abstract}
        
        
    \end{frontmatter}
    
    \section{Introduction}
    \label{s:introduction}
    
    Lung cancer is the leading cause of cancer death for both females and males \citep{siegel2020cancer}. Lung adenocarcinoma (ADC) is heterogeneous in morphological features, highly volatile in prognosis, and compromises half of all lung cancer cases \citep{matsuda2015morphological,wang2018comprehensive}. While mortality rates have rapidly declined, lung cancer in 2017 has caused more deaths than breast, prostate, colorectal, and brain cancer combined \citep{siegel2020cancer}.
    
    With the advanced imaging technology, digital hematoxylin and eosin (H\&E)-stained pathology imaging of tumor tissue slides is becoming a routine clinical procedure for cancer diagnosis. This process produces massive imaging data that capture histological details in high spatial resolution. Recent developments in deep-learning methods have enabled us to automatically detect and characterize the tumor regions in pathology images on a large scale \citep{wang2018comprehensive}. Through these developments, \citet{wang2018comprehensive} discovered a relationship between tumor shape and patient survival outcome in lung cancer patients. Although numerous studies have associated morphological features with patient prognosis \citep{wang2014novel,yu2016predicting,luo2017comprehensive}, this study was the first to incorporate shape into a prognostic model. Furthermore, this association reflects multiple studies in brain and breast cancer \citep{kilday1993classifying,pohlman1996quantitative}, though further systemic research with additional techniques to represent and quantify tumor shape is necessary.
    
    Existing approaches to describe the shape and boundary of regions have been the objective of shape analysis \citep{wirth2001shape,yang2008survey}. Shape representations characterize and visualize regions, while shape descriptors quantify the regions themselves. These shape representations, either one or two-dimensional, involve the regions or their contours. From these shape representations, we can compute shape descriptors to quantify the shape, geometry, and topology of the original regions. In general, shape descriptors should be translation, rotation, and scale-invariant \citep{yang2008survey}. Additionally, they should be application-dependent with a low computational complexity \citep{zhang2004review}.
    
\begin{figure}[H]
    	\centering
    	\includegraphics[width=0.9\linewidth]{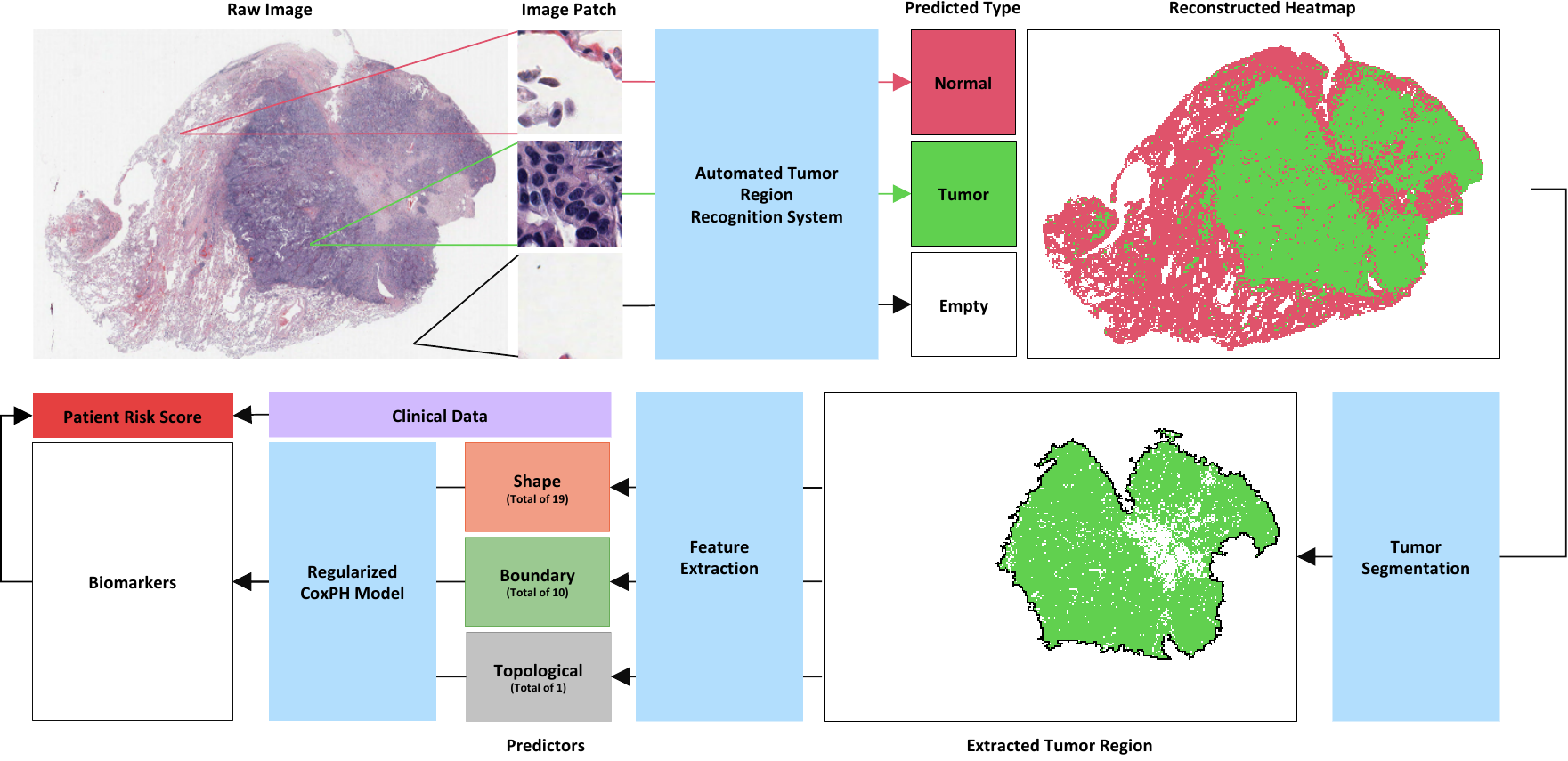}
    	\caption{Analysis pipeline with segmentation procedure and survival analysis.}
    	\label{f:workflow}
    \end{figure}

    In this paper, we develop a segmentation procedure and analysis pipeline to study the relationship between tumor shape and patient survival outcome. The analysis pipeline, based on \citet{wang2018comprehensive}, can be summarized in four parts: image pre-processing, feature extraction, survival analysis, and predictive performance. Image pre-processing relies on an automated tumor recognition system to convert raw pathology images into machine-readable representations; the results lead to a feature extraction process that computes shape,  geometric, and topological features from the extracted tumors in the pathology images. A regularized Cox proportional-hazards (CoxPH) model, dependent on the shape-based features, is used as an objective prognostic method. The evaluation of its predictive performance, adjusted for clinical variables, concludes the analysis pipeline. We summarize these steps in Figure~\ref{f:workflow}.

    We organize the remaining of this paper as follows. Section~\ref{s:segmentation} introduces the segmentation procedure that allows us to extract tumors from pathology images. Section~\ref{s:shape} presents the shape representations used to represent tumors and compute shape-based features; these features being defined in Section~\ref{s:feature}. In section~\ref{s:analysis}, we evaluate the prognostic performance of these features, while developing a shape-based prognostic model that varies from the one presented in \citet{wang2018comprehensive}. We discuss the results and conclude the paper in Section~\ref{s:conclusion}.
    
    \section{Tumor Segmentation}
    \label{s:segmentation}
    
    We developed a processing procedure to extract tumors in raw pathology images. Section~\ref{sb:heatmap} introduces the matrix representation of a whole-slide image that is used by the procedure and referred to as the reconstructed heat map. The intermediate tissue segmentation step is outlined in Section~\ref{sb:tissue}, while the main tumor segmentation step is detailed in Section~\ref{sb:segmentation}.
    
    \subsection{Reconstructed Heat Map}
    \label{sb:heatmap}
    
\begin{figure}[H]
    	\centering
    	\includegraphics[width=0.75\linewidth]{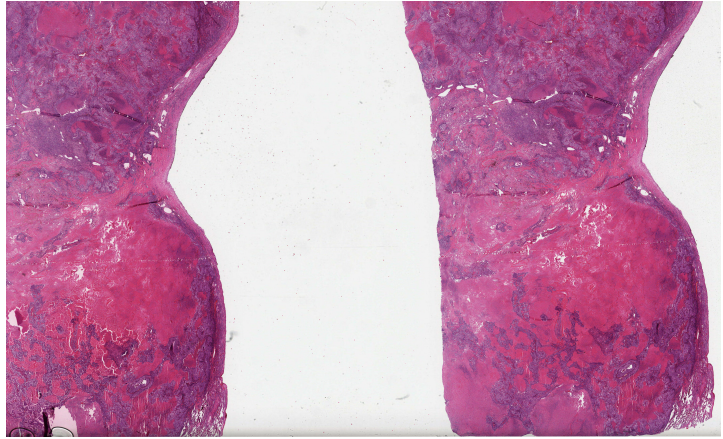}
    	\caption{An example of an H\&E stained pathology image from a lung cancer
    		patient in the TCGA cohort.}
    	\label{f:image}
    \end{figure}

\begin{figure}[H]
    	\centering
    	\includegraphics[width=0.9\linewidth]{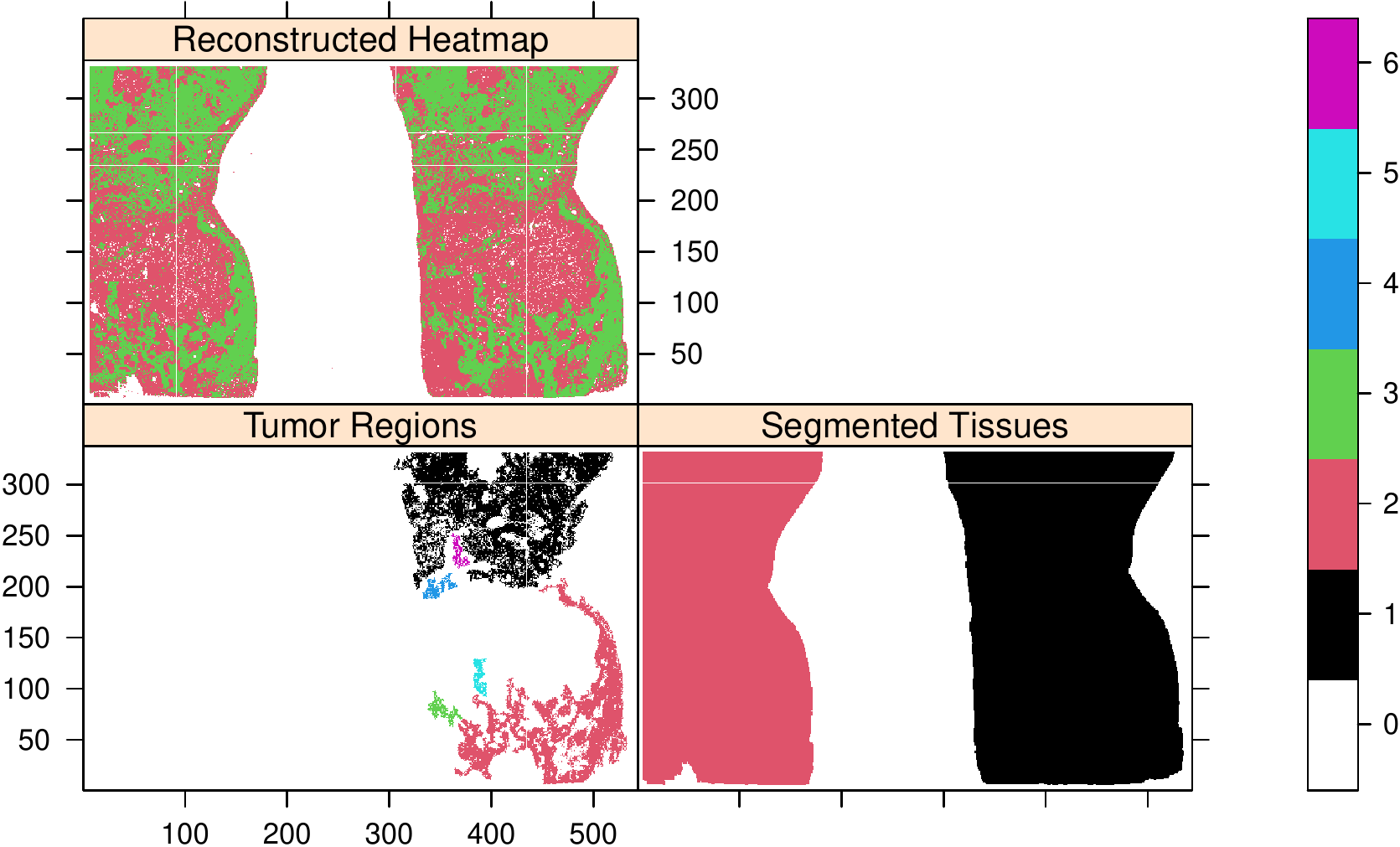}
    	\caption{Steps of tumor segmentation procedure.
    		Top-left figure shows the reconstructed heat map of the H\&E
    		stained pathology image.
    		Bottom-right and bottom-left figures show the identified tissue
    		regions and segmented tumors, respectively.
    		The key on the right shows the integer coding for each tumor.}
    	\label{f:segmentation}
    \end{figure}

    Automatic tumor recognition systems, such as the one developed in \citet{wang2018comprehensive}, create machine-readable representations of whole-slide images. A matrix results from the pathology image given to the recognition system or other imaging tools, where each entry corresponds to an empty region, non-malignant tissue, or tumor tissue in the raw sample. We refer to this matrix as a reconstructed heat map, which we can use to segment the tumor regions within the pathology image. For example, an automatic tumor recognition systems can be used with H\&E stained pathology images (see Figure~\ref{f:image}), where we can obtain a reconstructed heat map, as shown in Figure~\ref{f:segmentation} (top-left). We continue the processing procedure by segmenting the tissue regions.
    
    \subsection{Tissue Segmentation}
    \label{sb:tissue}
    
    Since a pathology image can contain multiple tissue samples, disconnected tissue regions are identified and segmented through standard morphological operations, usually based on a four-connectivity \citep{gonzalez2020digital}. We show this segmentation in Figure~\ref{f:segmentation} (bottom-left). Furthermore, we remove the tissue regions with an area less than one-fourth of the largest tissue region to remove the effects of small tissue samples \citep{wang2018comprehensive}. For each tissue region, we can finalize the processing procedure by segmenting the tumor regions.
    
    \subsection{Tumor Segmentation}
    \label{sb:segmentation}
    
    Individual tumor regions, which also contain empty and non-malignant categories within them, were extracted from each tissue region with a similar procedure as the tissue segmentation in Section~\ref{sb:tissue}. Similarly, to reduce the influence of small tumors, we disregard tumor regions with an area less than $150$ and consider them as ``islands'' to analyze, separately. We create two separate segmented matrices from the tumor regions and the non-malignant categories, respectively; the non-malignant categories make up the holes within the tumors. An integer coding is used to differentiate each tumor and their respective holes. We show the segmented tumor regions from the continued example in Figure~\ref{f:segmentation} (bottom-left).
    
    \section{Shape Representations}
    \label{s:shape}

\begin{figure}[H]
    	\centering
    \includegraphics[width=0.75\linewidth]{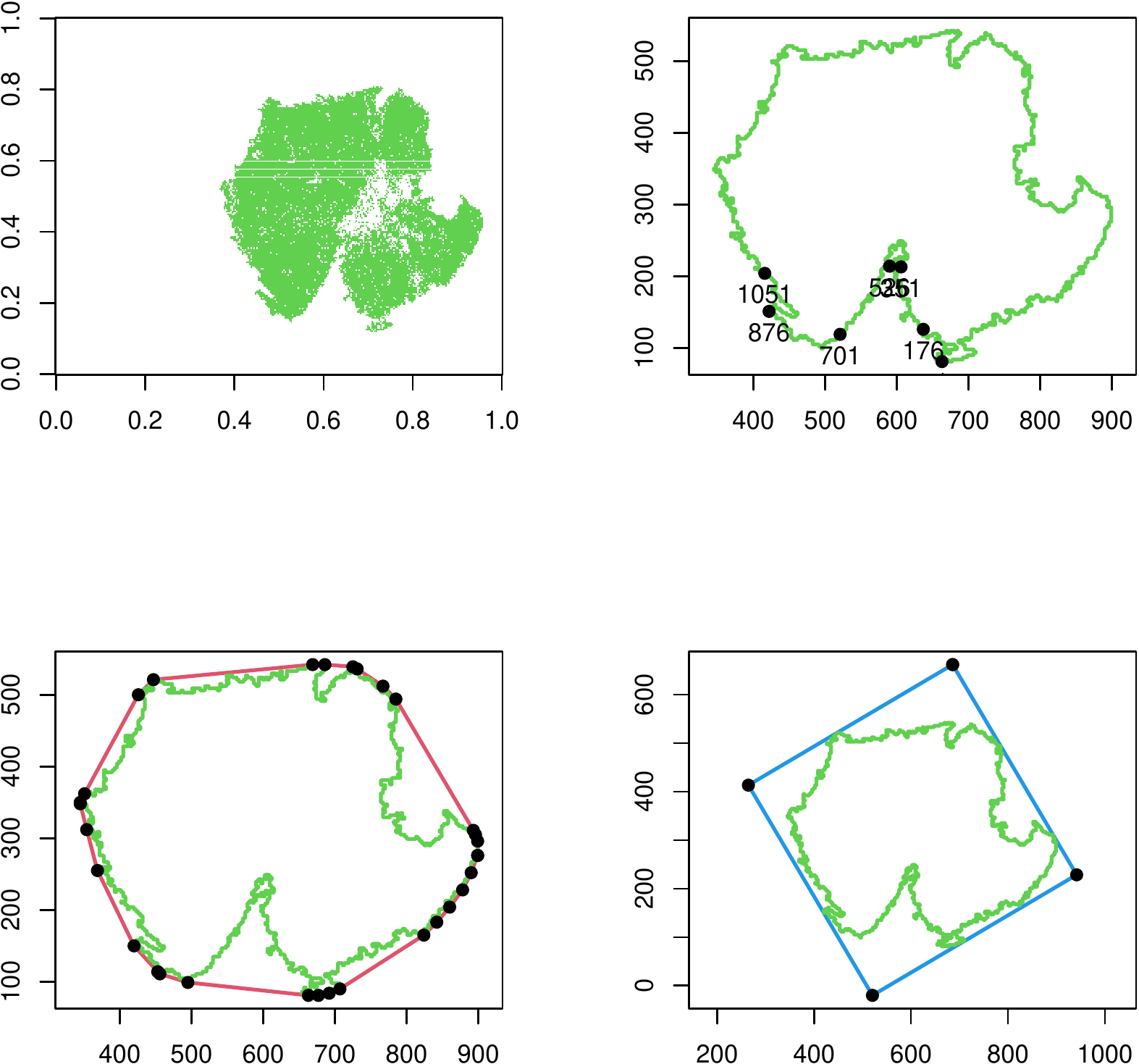}
    	\caption{Examples for the $2$-dimensional shape representations, based on a
    		tumor region in the NLST cohort.
    		Top two make up the primary representations: region matrix and polygon chain, respectively.
    		Bottom two make up the derived representations: convex hull and
    		minimum bounding box, respectively.
    		We overlay the  polygon chain on the derived representations.}
    	\label{f:2dshapes}
    \end{figure}

\begin{figure}[H]
    	\centering
    	\includegraphics[width=0.75\linewidth]{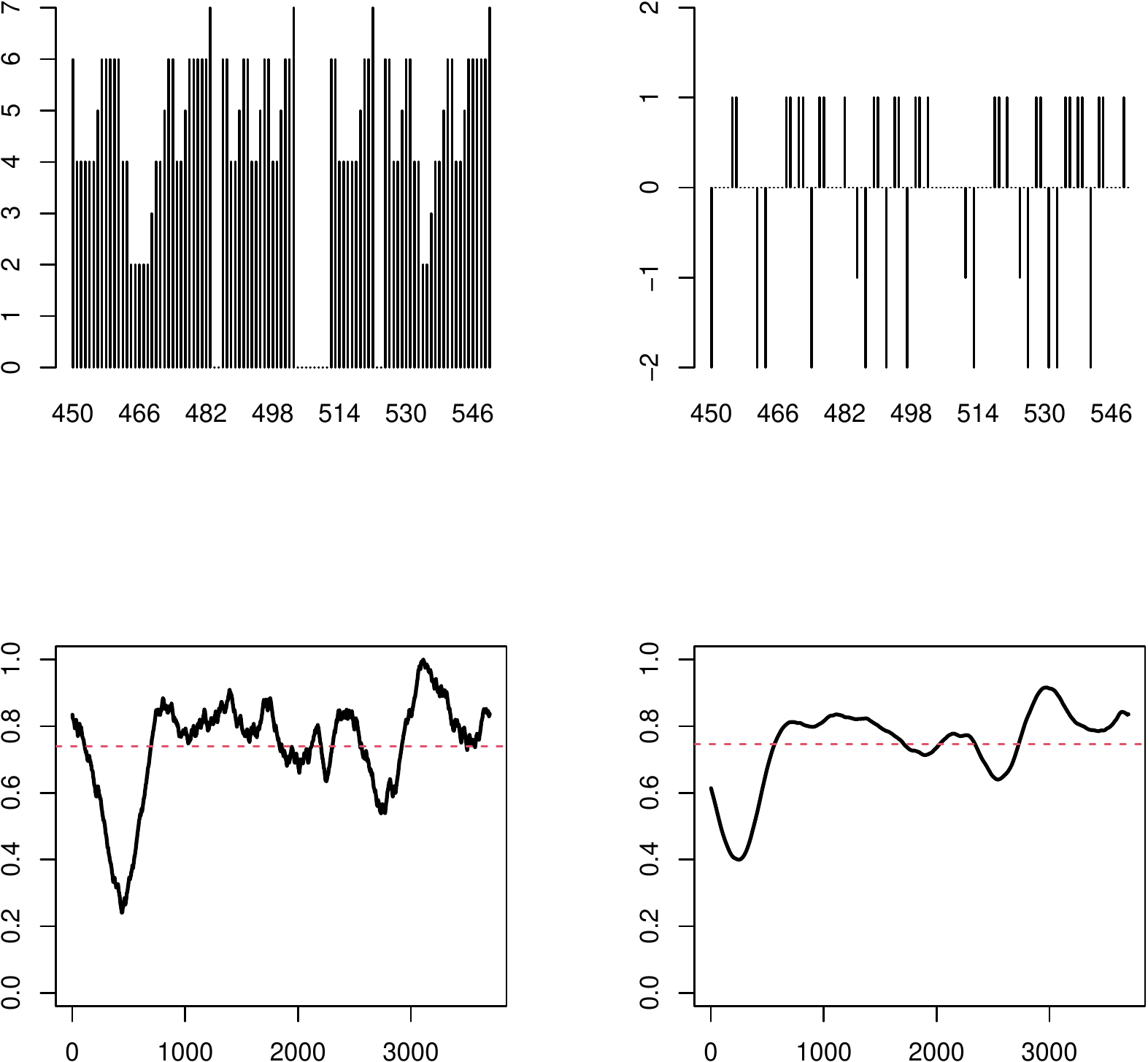}
    	\caption{Examples for the $1$-dimensional shape representations, based on
    		the tumor region in Figure~\ref{f:2dshapes}.
    		All four make up the derived shape representations: chain code,
    		curvature chain code, radial lengths, smoothed radial lengths,
    		respectively.
    		We overlay the mean for the two bottom representations.}
    	\label{f:1dshapes}
    \end{figure}

    From the segmented tumor regions, detailed in Section~\ref{sb:segmentation}, we can represent tumors with various one and two-dimensional techniques. These representation techniques should have invariance properties (rotation, scaling, translation), low computational complexity, and being application-independent \citep{yang2008survey}. Section~\ref{sb:region} and the beginning of Section~\ref{sb:polygon} make up the primary shape representations, while the remaining sections make up the derived shape representations.\footnote{The derived shape representations depend on the primary representations.} A visualization of these are shown in Figures~\ref{f:2dshapes}~and~\ref{f:1dshapes}, while Table~\ref{t:notation} summarizes the introduced notation. We begin by introducing the binary matrix, which is used to represent individual tumors in the segmented matrix.
    \begin{table}[H]
    	\centering
    	\footnotesize
    	\caption{Summary of original and derived shape representations.}
    	\begin{tabular}{l>{$}l<{$}>{$}l<{$}}
    		\toprule
    		
    		Name & \text{Notation} & \text{Support} \\ \midrule
    		
    		Region Matrix & \RG_{W \times L} & \iRG \in \set{0, 1} \\
    		
    		Polygon Chain & \PC_{(n+1) \times 2} & \iPC = (x_i, y_i) \in \N^2 \\
    		
    		Convex Hull Chain & \CH & \iCH \in \PC \\
    		
    		Bounding Box Chain & \BB & \iBB \in \R^2 \\
    		
    		Chain Code & \CC = \bracket[1 \times n]{\iCC} & \iCC \in \N \cap [0, 7] \\
    		
    		Curvature Chain Code & \K = \bracket[1 \times n]{\iK} & \iK \in \N \cap [-2, 2] \\
    		
    		Radial Lengths & \RL = \bracket[1 \times n]{\iRL} & \iRL \in \R^+ \\
    		
    		Normalized Radial Lengths & \NR = \bracket[1 \times n]{\iNR} & \iNR \in [0, 1] \\
    		
    		Smoothed Radial Lengths & \SR = \bracket[1 \times n]{\iSR} & \iSR \in [0, 1] \\
    		
    		\bottomrule
    	\end{tabular}
    	\label{t:notation}
    \end{table}
    
    \subsection{Binary Matrix}
    \label{sb:region}
    
    Let $\RG_{W \times L}$ be a binary matrix representing an arbitrary $W$-by-$L$
    image, containing a $4$-connected region, where the foreground and background
    are composed of ones and zeros, respectively.
    Additionally, we can represent each pixel in the image as a point in a
    $2$-dimensional discrete plane, that is, each entry $\iRG \in \RG$ can be
    denoted as a point $(l, w) \in \N^2$.
    Furthermore, to differentiate between foreground and background points,
    let $\I_R \colon \N^2 \to \set{0, 1}$ be the indicator function for an image
    matrix given by
    \[
        \I_R(l, w) = \begin{cases}
            1 & (l, w) \text{ is a foreground pixel, } \\
            0 & (l, w) \text{ is a background pixel. }
        \end{cases}
    \]
    The indicator function $\I_R$ and distribution of points $(l, w)$'s will be used
    to recreate the region's contour in a two dimensional Cartesian plane, known as the
    polygon chain.
    
    \subsection{Polygon Chain}
    \label{sb:polygon}
    
    We can find the entries in the image matrix that make up the region's boundary
    to, subsequently, obtain the equivalent points by using the Moore-Neighbor
    tracing algorithm, modified by Jacob's stopping
    criteria~\cite{gonzalez2020digital}.
    The algorithm has three arguments:
    \begin{enumerate}
        \item starting boundary point,
        \item direction to traverse the boundary (clockwise or counter-clockwise), and
        \item the pixel connectivity.
    \end{enumerate}
    
    \noindent
    Clearly, since the image matrix contains a $4$-connected component, the pixel
    connectivity must be four.
    Furthermore, in our case, the boundary shall be traced in a clockwise direction.
    For the starting boundary point, we shall choose the point that represents the
    lowest left-most entry in the region.
    That is, let
    \[
        S = \set{ (l_i, w_i) \colon \I_R(l_i, w_i) = 1 \land w_i = \min[1 \leq k \leq W] w_k }
    \]
    be the collection of points that make up the region and are located in the
    lowest $y$-coordinate such that
    \[
        (l_1, w_1) = \begin{cases}
            S_1         & \abs{S} = 1, \\
            \min[l_i] S & \text{ otherwise. }
            \end{cases}
    \]
    Applying the Moore-Neighbor tracing algorithm to the region matrix, results in
    the points that make up the boundary of the region, specifically, where the
    boundary begins at the lowest left-most area of the region and traverse
    through the boundary in a clockwise direction.
    
    From the boundary points, we can create a sequence of points, known as the
    closed polygon chain, that represents the boundary of the region by creating
    a closed and simple polygon.
    Let $N = \sum_{l,w} \I_R(l, w)$ be the total number of points that make up the
    region and $\PC_{(n+1) \times 2}$ be the collection of points that make up the
    closed polygon chain of the region such that
    \begin{enumerate}
        \item $n \leq N$ is the number of boundary points,
        \item $\iPC = (x_i, y_i) \in \N^2$, for $i = 1, \ldots, (n+1)$, and
        \item $(x_1, y_1) = (x_{n+1}, y_{n+1}).$
    \end{enumerate}
    
    \noindent
    Through the polygon chain, we can derive further two- and one-dimensional
    shape representations that can be used to compute specific descriptors.
    
    \subsubsection{Convex Hull Chain}
    \label{sbb:convex}
    
    The convex hull of an arbitrary shape is defined as the smallest convex
    polygon that encloses the shape.
    Therefore, we can introduce the collection of points which create a closed and
    simple polygon, known as the convex hull chain and denoted by $\CH$, such that
    \begin{enumerate}
        \item the convex polygon created by $\CH$ encloses the shape,
        \item the number of points in $\CH$ is less than or equal to $n$, and
        \item each point $\iCH \in \PC$.
    \end{enumerate}
    
    \noindent
    Since the polygon chain forms a closed and simple polygon, we can use the
    linear time algorithm introduced in \citet{lee1983finding} to obtain the
    points $\iPC \in \PC$ that make up the convex hull of the shape to,
    subsequently, create the convex hull chain using the directional and starting
    point requirements in Section~\ref{sb:polygon}.
    
    \subsubsection{Bounding Box Chain}
    \label{sbb:box}
    
    The minimum area rectangle, or MAR, encloses an arbitrary shape with the
    smallest area possible~\cite{freeman1975determining}.
    Let the minimum bounding box chain, denoted as $\BB$, be the $4 \times 2$
    vector that makes up of the collection of points in the minimum area rectangle.
    To determine these points, we make use of the theorem
    in \citet{freeman1975determining} which states that the minimum area rectangle
    has a side collinear with the side of the convex polygon it encloses.
    
    That is, to obtain the MAR of the shape formed by the polygon chain, we need
    the convex hull chain.
    Additionally, at least one side of the convex hull, formed by two
    points $\iCH[i]$ and $\iCH[i+1]$, will align with one side of the MAR.
    As a result, the linear-time algorithm presented
    in \citet{toussaint1983solving} can be applied to $\CH$ to construct $\BB$.
    This algorithm obtains the points $\iBB \in \BB$ and decreases the running
    time of the original algorithm in \citet{freeman1975determining}
    from $\bigO(n^2)$ to $\bigO(n)$ by using rotating
    calipers~\cite{toussaint1983solving}.
    
    \subsection{Chain Code}
    \label{sb:code}
    
    The slope of a shape's contour can be approximated by the directional changes
    between two consecutive boundary points.
    These directional changes can be encoded to, essentially, assign a
    number (from $0$ to $7$) to each possible relative direction resulting in an
    encoding list, each element known as the chain code, that provides a
    compact representation of the shape's
    contour~\cite{wirth2001shape,lecture8slides}.
    Let $\CC$ be a $1 \times n$ vector representing the chain codes of the polygon
    chain where each entry $\iCC \in \N \cap [0, 7]$ corresponds to a direction in
    the $8$-way split of the unit circle and determined by a series of steps.
    
    First, we determine the angle between the vector composed of the difference
    between the two consecutive points and the $x$-axis, that is,
    let $\theta_i = \arctantwo(\mathbf{d}_i)$ be the resulting angle
    where $\mathbf{d}_i = \iPC[i] - \iPC[i+1]$ is the difference.
    Since $\theta_i \in [-\pi, \pi]$, we have to transform the angle to
    \[
        \hat{\theta}_i = \begin{cases}
            \theta_i        & \theta_i \geq 0, \\
            \theta_i + 2\pi & \theta_i < 0,
        \end{cases}
    \]
    such that $\hat{\theta}_i \in [0, 2\pi)$.
    As a result, we can now determine the corresponding chain code
    \[
        \iCC = \round{\frac{\hat{\theta}_i}{\pi/4}}.
    \]
    
    \noindent
    Clearly, we can see that this procedure splits the unit circle into eight
    equal parts.
    Additionally, if the directional change does not exactly align within the eight
    splits, then the rounding operator $\round{\ }$ will approximate the chain code
    to the nearest integer.
    
    \subsubsection{Curvature Chain Code}
    \label{sbb:curvature}
    
    Let $\K$ be a $1 \times n$ vector representing the curvature chain code such
    that each entry is formed from a transformation of the difference between two
    consecutive chain codes, that is, let $\hat{d}_i = \iCC[i] - \iCC[i+1]$ and
    \[
        \iK = \begin{cases}
            \hat{d}_i + 7  & \hat{d}_i < 2, \\
            \hat{d}_i - 7  & \hat{d}_i > 2, \\
            \hat{d}_i      & \text{otherwise. }
        \end{cases}
    \]
    
    \noindent
    This simple chain code derivation estimates the curvature and contains
    information on the convexity of the shape~\cite{wirth2001shape}.
    
    \subsection{Radial Lengths}
    \label{sb:radial}
    
    Let $\RL$ be a $1 \times n$ vector of radial lengths, that is, each
    entry $\iRL = \norm{\iPC - \cent}$, for $i = 1, \ldots, n$, is the Euclidean
    distance from the boundary point to the shape's
    centroid (See Appendix~\ref{a:centroid}).
    Clearly, the radial lengths are not
    scale-invariant (as the Euclidean distance is not).
    Therefore, to properly analyze the structure of $\RL$, the individual radial
    lengths must be normalized.
    
    Let $r_{(n)}$ be the maximum radial length in $\RL$ such that we can introduce
    a $1 \times n$ vector of normalized radial lengths, denoted as $\NR$, where
    each entry $\iNR = \iRL\;/\;r_{(n)}$, $i = 1, \ldots, n$.
    By normalizing the radial lengths, we have obtain a $1$-dimensional signal
    that is scale-invariant and which we can use to analyze the fine details of the
    shape's contour~\cite{wirth2001shape}.
    
    \subsubsection{Smoothed Radial Lengths}
    \label{sbb:smoothed}
    
    While the normalized radial lengths make up a scale-invariant representation
    of the shape's contour, this can still produce extra noise that impacts
    shape descriptors.
    We can remove any noise in the $1$-dimensional signal of the shape by smoothing
    the normalized radial lengths; specifically, we apply a mean filter with a
    size equal to $10\%$ of the shape's perimeter to the normalized
    radial lengths~\cite{pohlman1996quantitative}.
    Let the window size $w$ be $10\%$ of the shape's perimeter and $\SR$ be
    a $1 \times n$ vector representing the smoothed radial lengths such that
    each entry
    \[
        \iSR = \frac{1}{w} \sum_{j=i}^{k} \I(j \leq n) \cdot \iNR[j], \quad
            k = i+w-1, \quad i = 1, \ldots, n.
    \]
    
    \section{Feature Extraction}
    \label{s:feature}

    From the shape representations in Section~\ref{s:shape}, we can compute 
    shape-based features, known as shape descriptors.
    These features allow us to quantify the shape, geometry, and topology of 
    a tumor and are categorized into either shape, boundary, or topological 
    features. 
    
    The continuation of the processing procedure in Section~\ref{s:segmentation},
    relies on converting the segmented tumors in Section~\ref{sb:segmentation}
    to individual binary matrices. 
    Furthermore, The polygon chain for the contour of each tumor is created 
    and the derived shape representations are also obtained.
    From these results, we can compute the tumor-level features of each category.
    
\begin{figure}[H]
    	\centering
    	\includegraphics[width=1.0\linewidth]{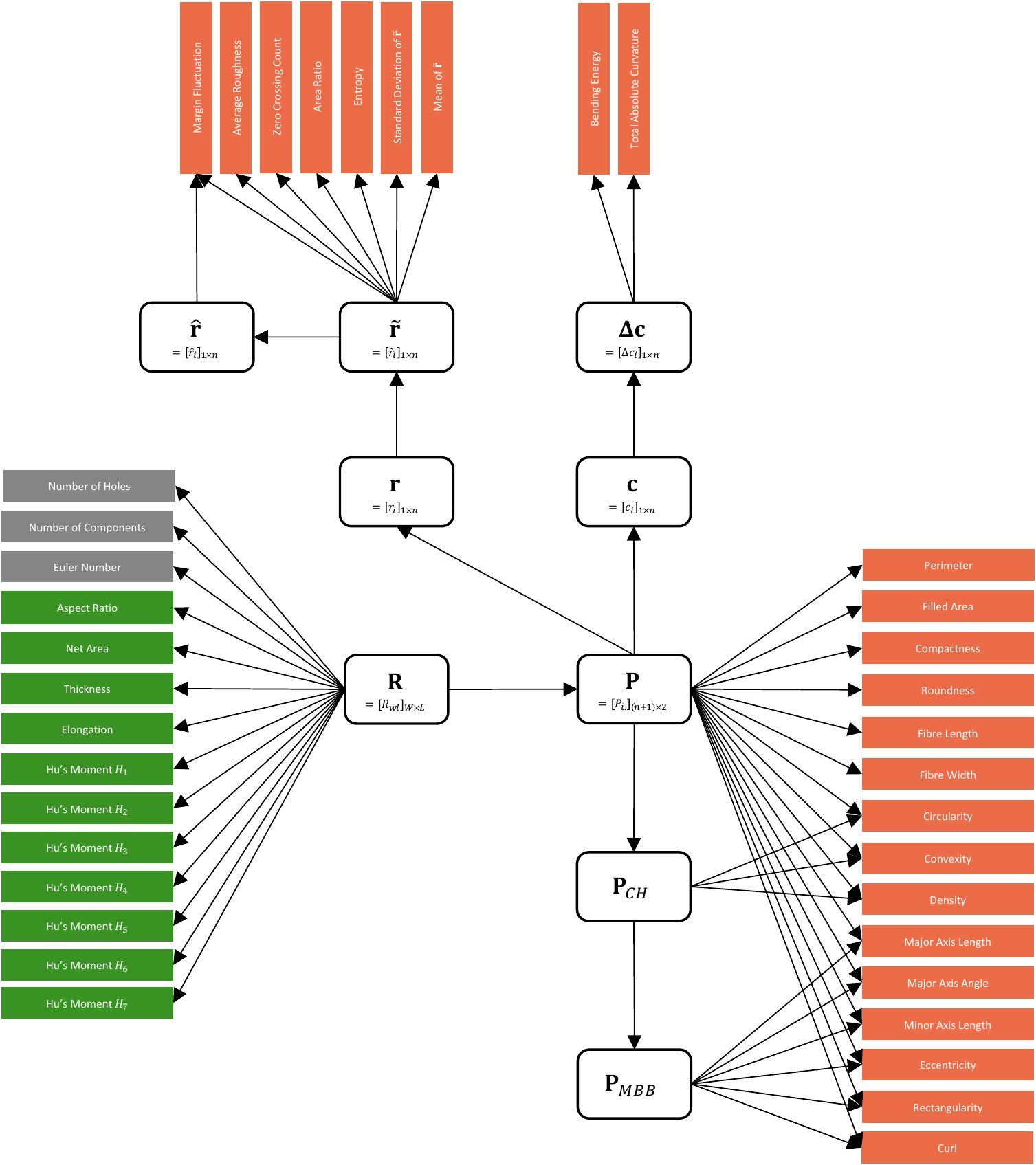}
    	\caption{Dependencies of shape descriptors and representations.
    		Colored boxes refer to the categories of the shape descriptors.
    		Green denotes geometrical, grey topological, and orange shape.}
    	\label{f:reps}
    \end{figure}

    These categories and their respective features are introduced, below, while
    Table~\ref{t:features} shows an example of some of the features
    extracted from the tumor regions in Figure~\ref{f:segmentation} (bottom-left). 
    In Section~\ref{sb:shape}, we introduce the shape features while in 
    Sections~\ref{sb:boundary}~and~\ref{sb:topological} we introduce the
    boundary and topological features, respectively. 
    The dependencies for these features are summarized in Figure~\ref{f:reps}.
    \begin{table}[H]
    	\caption{Computed features from the tumor regions in
    		Figure~\ref{f:segmentation}.
    		Each row corresponds to a tumor where the Tumor ID is based on
    		the integer coding found on the key of
    		Figure~\ref{f:segmentation}.}
    	
    	\footnotesize
    	
            \begin{tabular}{crrrrrrrr}
                \toprule
                Tumor & Thick- & \# of  & Perimeter &  & Major Axis  & Minor Axis  & Fibre  & Fibre \\
                 ID & ness &Holes & Perimeter & Area &  Length &  Length &  Length &  Width\\
                \midrule
                1 & 37 & 873 & 5801.6652 & 159203.0 & 644.0000 & 404.00000 & 55.961409 & 2844.8712\\
                2 & 24 & 328 & 6928.8683 & 96094.5 & 608.2032 & 501.90189 & 27.963134 & 3436.4710\\
                3 & 5 & 26 & 639.1371 & 3242.0 & 117.2818 & 67.34740 & 10.489217 & 309.0793\\
                4 & 5 & 24 & 667.2376 & 2821.5 & 106.7748 & 63.32211 & 8.683261 & 324.9355\\
                5 & 4 & 17 & 470.8528 & 2458.0 & 113.0000 & 38.00000 & 10.949922 & 224.4765\\
                6 & 3 & 19 & 509.9239 & 2016.5 & 105.1036 & 47.72370 & 8.170879 & 246.7911\\
                \bottomrule
            \end{tabular}
    	
    	\label{t:features}
    \end{table}
    
    \subsection{Shape Features}
    \label{sb:shape}
    
    The shape features characterize the geometry and shape of tumor, relying on
    the $2$-dimensional shape representations, that is, the binary matrix $\RG$,
    polygon chain $\PC$, convex hull chain $\CH$, and bounding box chain $\BB$.
    We begin by introducing the region moments, features derived from regular 
    statistical moments.
    
    \subsubsection{Region Moments}
    
    Since we can view the entries in the region matrix as points in a
    $2$-dimensional discrete plane, we can view the points as a statistical
    distribution in a $2$-dimensional space~\cite{lecture8slides}.
    From this statistical distribution, we can calculate the concrete physical
    properties, known as statistical moments, of the region~\cite{lecture8slides}.
    
    We start with the low-order moments, denoted as $m_{pq}$ and known as the
    ordinary moments of order $p$ and $q$, given by
    \[
        m_{pq} = \sum_{l,w} \I_R(l, w) \cdot l^p w^q
    \]
    and identical to standard statistical moments; that is, $m_{pq}$ takes
    the $p^\text{th}$ moment in the $l$-direction and the $q^\text{th}$ moment in
    the $w$-direction.
    
    Clearly, ordinary moments are not translation-invariant as the values depend
    on the $lw$-coordinates.
    Therefore, central moments, denoted as $\mu_{pq}$, measure characteristic
    properties with respect to the region's centroid $(\bar{l}, \bar{w})$ which
    result in translation-invariant features.
    The central moments are, hence, given by
    \[
        \mu_{pq} = \sum_{l,w} \I_R(l, w) \cdot (l - \bar{l})^p \cdot (w - \bar{w})^q
    \]
    where the centroid $(\bar{l}, \bar{w})$ is determined by
    \[
        \bar{l} = \frac{m_{10}}{m_{00}} \quad \text{and} \quad \bar{w} = \frac{m_{01}}{m_{00}}.
    \]
    
    \noindent
    Although central moments are translation-invariant, they are impacted by the
    area of the region and are, thus, not scale-invariant.
    
    Normalizing the central moments by some uniform factor allows the computation
    of translation- and scale-invariant features~\cite{lecture8slides}. This
    normalization procedure yields the normalized central moments, denoted
    as $\eta_{pq}$, given by
    \[
        \eta_{pq} = \mu_{pq} \cdot \left( \frac{1}{\mu_{00}} \right)^{(p+q+2)/2}
    \]
    such that $(p+q) \geq 2$. With the normalized central moments, we can derive
    the linear combinations known as Hu's moments, denoted as $H_i$, which are
    invariant to translation, rotation, and scaling~\cite{lecture8slides}.
    Although Hu's moments are linear, their formulation cannot be generalized.
    Therefore, we only give the first three of Hu's moments given by
    \begin{align*}
        H_1 &= \eta_{20} + \eta_{02} \\
        H_2 &= (\eta_{20} + \eta_{02})^2 + 4 \cdot \eta_{11}^2 \\
        H_3 &= (\eta_{30} - 3 \cdot \eta_{12})^2 + (3 \cdot \eta_{21} - \eta_{03})^2 \\
            &\vdots
    \end{align*}
    
    \subsubsection{Aspect Ratio}

    The aspect ratio is the relationship between the dimensions of the
    image and is given by
    \[
        \text{AspectRatio}(\RG) = \frac{W}{L},
    \]
    where $W$ and $L$ are the width and length of the image,
    respectively~\cite{wirth2001shape}.
    Unlike other descriptors, the aspect ratio does not depend on the shape of
    the region within the image.
    
    \subsubsection{Net Area}
    
    The net area is the number of pixels that make up the region:
    \[
        \text{Area}^\text{Net}(\RG) = \sum_{l,w} \iRG = \abs{\RG}.
    \]
    
    \subsubsection{Thickness}
    
    The thickness of a region, without holes, is determined by the number of of
    erosion steps until the region disappears:
    \[
        \text{Thickness}(\RG) = \abs{E} = d,
    \]
    where $E \coloneqq \set{\RG_i}$ is the sequence of eroded region matrices
    such that $\RG_i = \text{Erode}(\RG_{i-1})$, $\RG_1 = \RG$,
    and $\abs{\RG_d} = 0$, for $i = 1, \ldots, d$~\cite{wirth2001shape}.
    
    \subsubsection{Elongation}
    
    The elongation of a region is the relationship between the area and the
    squared-thickness, given by
    \[
        \text{Elongation}(\RG) =
            \frac{\text{Area}^\text{Net}(\RG)}{2 \cdot (\text{Thickness}(\RG))^2};
    \]
    we note that this descriptor becomes distorted as the curvature of the
    region increases~\cite{wirth2001shape}.
    
    \subsubsection{Perimeter}

    The perimeter is the length of the region's contour, an approximation to
    number of pixels that make up the region:
    \[
        \text{Perimeter}(\PC) = \sum_{i=1}^n \norm{\iPC - \iPC[i+1]}.
    \]
    
    \subsubsection{Filled Area}
    
    The filled area is the approximate area of the polygon that represents the
    region, estimated using Gauss's area formula:
    \[
        \text{Area}^\text{Filled}(\PC) \approx \frac{1}{2}
            \abs{ \sum_{i=1}^n (x_i y_{i+1} - x_{i+1} y_i) }
    \]
    This descriptor is also known as the area of connected region, that is, the
    area of the region without holes~\cite{lecture8slides}.
    
    \subsubsection{Compactness}
    
    The compactness of a region is a dimensionless quantity that increases as
    the irregularity of the shape increases, given by the relationship between
    perimeter and area~\cite{castleman1996digital}. That is, the formulae is
    given by
    \[
        \text{Compactness}(\PC) =
            \frac{(\text{Perimeter}(\PC))^2}{\text{Area}^\text{Filled}(\PC)}.
    \]
    
    \subsubsection{Roundness}
    
    The roundness of a region is the compactness normalized against a filled
    circle, that is, the relationship between the area of the region and the
    area of a circle with the same perimeter~\cite{wirth2001shape}.
    The formulae is given by
    \[
        \text{Roundness}(\PC) = 4\pi \cdot
            \frac{\text{Area}^\text{Filled}(\PC)}{(\text{Perimeter}(\PC))^2}
    \]
    
    \subsubsection{Fibre Length and Width}
    
    The fibre length and width are, clearly, another form of measurements for a shape's
    length and width, respectively. The fibre length is given by
    \[
        \text{Fibre}^\text{Length}(\PC) =
            \frac{\text{Perimeter}(\PC) -
                \sqrt{(\text{Perimeter}(\PC))^2 -
                    16 \cdot \text{Area}^\text{Filled}(\PC)}}{4}
    \]
    and the fibre width by
    \[
        \text{Fibre}^\text{Width}(\PC) = \frac{\text{Area}^\text{Filled}(\PC)}
            {\text{Fibre}^\text{Length}(\PC)}.
    \]
    
    \subsubsection{Circularity}
    
    The circularity of a region is the relationship between the area of the
    region and the area of a circle with same convex
    perimeter:
    \[
        \text{Circularity}(\PC, \CH) = 4\pi \cdot
            \frac{\text{Area}^\text{Filled}(\PC)}{(\text{Perimeter}(\CH))^2}.
    \]
    This descriptor decreases as the shapes becomes less circular, but is
    relatively insensitive to irregular boundaries~\cite{wirth2001shape}.
    
    \subsubsection{Convexity}
    
    The convexity of a region is the relationship between the perimeter of the
    region's convex hull and the perimeter of the region:
    \[
        \text{Convexity}(\PC, \CH) =
            \frac{\text{Perimeter}(\CH)}{\text{Perimeter}(\PC)}.
    \]
    Therefore, this descriptor quantifies how much the shape differs from a
    convex shape as well as accounting for any irregularities along the
    boundary of the shape~\cite{wirth2001shape}.
    
    \subsubsection{Density}
    
    The density of a region is a measurement of solidity, obtained from the
    ratio of the area of the region to the area of the region's
    convex hull~\cite{wirth2001shape}.
    That is,
    \[
        \text{Density}(\PC, \CH) = \frac{\text{Area}^\text{Filled}(\PC)}
            {\text{Area}^\text{Filled}(\CH)}.
    \]
    
    \subsubsection{Major Axis Length}
    
    The major axis length is the length of the line segment that runs along
    the longest part of the region and is, clearly, an approximate measurement
    of the region's length~\cite{wirth2001shape,lecture8slides}.
    The formulae is given by
    \[
        \text{MajorAxisLength}(\BB) = \norm{b_l},
    \]
    where $l = \argmax[1 \leq i \leq 4]\norm{b_i}$ and $b_i = \iBB - \iBB[i+1]$.
    
    \subsubsection{Major Axis Angle}
    
    The major axis angle describes the direction of the major axis, relative to
    the $x$-axis, and is given by
    \[
        \text{MajorAxisAngle}(\BB) = \arctantwo(b_k).
    \]
    
    \subsubsection{Minor Axis Length}
    
    Opposite to the major axis length, the minor axis length is the length of the line segment that runs along the shortest part of the region~\cite{lecture8slides}.
    Clearly, an approximate measurement of the region's
    width~\cite{wirth2001shape}.
    The formulae is given by
    \[
        \text{MinorAxisLength}(\BB) = \norm{b_s},
    \]
    where $s = \argmin[1 \leq i \leq 4]\norm{b_i}$ and $b_i = \iBB - \iBB[i+1]$.
    
    \subsubsection{Eccentricity}
    
    The eccentricity of a region is the relationship between the major and minor
    axis; this descriptor is another way to measure the elongation of a region
    and is given by
    \[
        \text{Eccentricity}(\BB) = \frac{\text{MajorAxisLength}(\BB)}
            {\text{MinorAxisLength}(\BB)}
    \]
    We notice that as the eccentricity approaches one, the shape of the region
    becomes less elongated and roughly square or circular~\cite{wirth2001shape}.
    
    \subsubsection{Rectangularity}
    
    The rectangularity of a region is the relationship between the area of a
    region and the area of the region's minimum bounding box, given by
    \[
        \text{Rectangularity}(\PC, \BB) = \frac{\text{Area}^\text{Filled}(\PC)}
            {\text{Area}^\text{Filled}(\BB)}.
    \]
    
    \subsubsection{Curl}
    
    The curl of a region quantifies how much its shape is ``curled up'',
    given by
    \[
        \text{Curl}(\PC, \BB) = \frac{\text{MajorAxisLength}(\BB)}
            {\text{Fibre}^\text{Length}(\PC)}.
    \]
    We can see that as the descriptor decreases, the degree to which the region is
    ``curled up'' increases~\cite{wirth2001shape}.
    
    \subsection{Boundary Features}
    \label{sb:boundary}
    
    The boundary features characterize the contour of the tumor, relying on the
    $1$-dimensional shape representations, that is, the chain code $\CC$, 
    curvature chain code $\K$, radial lengths $\NR$, smoothed radial lengths
    $\SR$. 
    We begin by introducing the contour-equivalent of the region moments: the
    boundary moments. 
    We rely on the radial lengths $\RL$, but \citet{yang2008survey} notes that 
    any $1$-dimensional representation works.
    
    \subsubsection{Boundary Moments}
    
    Using the radial lengths as a $1$-dimensional representation, we can reduce
    the dimension of the shape with boundary moments~\cite{yang2008survey,sonka2014image}.
    Similar to region moments, we can estimate both the $p^\text{th}$
    moment $m_p$ and central moment $\mu_p$ by
    \[
        m_p = \frac{1}{n} \sum_{i=1}^n (\iRL)^p \quad \text{and} \quad \mu_p = \frac{1}{n} \sum_{i=1}^n (\iRL - m_1)^p.
    \]
    
    \noindent
    Additionally, we also have two normalized moments given by
    \[
        \bar{m}_p = \frac{m_p}{(\mu_2)^{r/2}} \quad \text{and} \quad \bar{\mu}_p = \frac{\mu_p}{(\mu_2)^{r/2}}
    \]
    which are invariant to translation, rotation, and scaling.
    From these moments, less noise-sensitive descriptors can be obtained and are given by
    \[
        F_1 = \frac{(\mu_2)^{1/2}}{m_1}, \quad F_2 = \frac{\mu_3}{(\mu_2)^{3/2}}, \quad \text{and} \quad F_3 = \frac{\mu_4}{(\mu_2)^2}.
    \]
    
    \subsubsection{Bending Energy}
    
    The total bending energy of a region is, theoretically, the energy necessary
    to bend a rod the shape of the region~\cite{wirth2001shape}.
    As a result, the minimum value for this descriptor is $\frac{2\pi}{R}$ for
    a circle of radius $R$~\cite{wirth2001shape}.
    This robust descriptor is given by
    \[
        \text{E}_\K = \frac{1}{n} \sum_{i=1}^n (\iK)^2.
    \]
    
    \subsubsection{Total Absolute Curvature}
    
    The total absolute of a curvature for a region is given by
    \[
        \K_\text{total} = \frac{1}{n} \sum_{i=1}^n \abs{\iK},
    \]
    where the minimum is obtained for all convex shapes~\cite{wirth2001shape}.
    
    \subsubsection{Radial Mean and Standard Deviation}
    
    The average of the normalized radial lengths is given by
    \[
        \bar{r} = \frac{1}{n} \sum_{i=1}^n \iNR
    \]
    and can measure macroscopic changes in the shape's contour.
    Similarly, the standard deviation is given by
    \[
        s_r = \sqrt{ \frac{1}{n} \sum_{i=1}^n (\iNR - \bar{r})^2 }
    \]
    and can indicate fine contour changes.
    
    \subsubsection{Circularity}
    
    Another derivation for the circularity of a region, using the normalized
    radial lengths, is given by
    \[
        \text{Circularity}(\NR) = \frac{\bar{r}}{s_r},
    \]
    where $\bar{r}$ and $s_r$ are the mean and standard deviation of the radial
    lengths, respectively.
    This derivation increases as the shape becomes more
    circular~\cite{haralick1974measure}.
    
    \subsubsection{Entropy}
    
    The entropy relies on the $K$-bin histogram of the radial lengths and is a
    probabilistic measure that quantifies how well the radial lengths can be
    estimated~\cite{kilday1993classifying,wirth2001shape}.
    The formulae is given by
    \[
        E(\NR) = -\sum_{k=1}^K p_k \log(p_k),
    \]
    where $p_k$ is the $k^{th}$ entry of the $K$-bin histogram\footnote{
        \citet{kilday1993classifying} relied on the $100$-bin probability
        histogram to compute the entropy of tumor regions.
    }.
    This descriptor incorporates the roundness and roughness of a
    shape~\cite{kilday1993classifying}.
    
    \subsubsection{Area Ratio}
    
    The area ratio measures how much of the shape is outside the circle with a
    radius defined by the radial mean, describing the macroscopic
    characteristics of the shape~\cite{kilday1993classifying}.
    The formulae is given by
    \[
        \text{Area}^\text{Ratio}(\NR) = \frac{1}{n \cdot \bar{r}}
            \sum_{i=1}^n \I(\iNR > \bar{r}) \cdot (\iNR - \bar{r}).
    \]
    
    \subsubsection{Zero-crossing Count}
    
    In conjunction with the area ratio, the zero-crossing count measures the
    number of times the shape crosses the contour of the circle with a radius
    defined by the radial mean. This descriptor captures the fine detail of
    the shape's contour and is given by
    \[
        \text{ZeroCrossing}(\NR) = \sum_{i=1}^n
            \I \Big\{ (\iNR - \bar{r}) \cdot ( \iNR[i+1] - \bar{r}) < 0 \Big\}.
    \]
    
    \subsubsection{Average Roughness Index}
    
    The average roughness index divides the contour of the shape into small
    segments of equal length, computing a roughness measure on each segment and
    averaging the results~\cite{kilday1993classifying}.
    That is,
    \[
        \text{Ave}_\text{R}(\NR) = \floor{\frac{l}{n}} \sum_{j=1}^{\floor{n/l}} R(j),
    \]
    where $l$ is the number of points in a fixed interval and
    \[
        R(j) = \sum_{k=j}^{l+j} \abs{\iNR[k] - \iNR[k+1]}, \quad
            j = 1, \ldots, \floor{\frac{n}{l}}.
    \]
    
    \subsubsection{Margin Fluctuation}
    
    The margin fluctuation is the standard deviation of the differences between
    the normalized radial lengths and their corresponding smoothed radial
    length, that is,
    \[
        \text{MarginFluctuation}(\NR, \SR) = \sqrt{
            \frac{1}{n} \sum_{i=1}^n (d_i - \bar{d})^2
        }
    \]
    where $d_i = \iNR - \iSR$ and
    \[
        \bar{d} = \frac{1}{n} \sum_{i=1} d_i.
    \]
    This descriptor was shown to be greater for spiculated shapes
    in \citet{giger1994computerized}.
    
    \subsection{Topological Features}
    \label{sb:topological}
    
    The topological features are not impacted by degradations or deformations in
    the shape~\cite{wirth2001shape}. 
    Therefore, they are invariant to rotation, scaling, and translation. 
    These makes them useful global shape descriptor, that can characterize the
    structure of a region~\cite{wirth2001shape,lecture8slides}.
    Features based on topological data analysis, have been a rising focus of 
    shape analysis. 
    Although, in this section, we introduce only three simple topological features.
    
    \subsubsection{Number of Holes}

    The number of holes within a region is a simple and robust measurement,
    given by
    \[
        N_H(\RG) =\max[k]\text{Segment}(\HR),
    \]
    where $\HR \coloneqq \text{FillHoles}(\RG) - \RG$ is the matrix that
    contains the holes.
    
    \subsubsection{Number of Connected Components}
    
    The number of connected components is, trivially, the number of regions in
    the image based on a $4$-connectivity or
    $8$-connectivity~\cite{gonzalez2020digital} and is given by
    \[
        N_R(\RG) = \max[k] \text{Segment}(\RG).
    \]
    
    \subsubsection{Euler Number}
    
    The Euler number is the number of connected components minus the number of
    holes:
    \[
        N_E(\RG) = N_R(\RG) - N_H(\RG).
    \]
    This simple descriptor is invariant to rotation, scaling,
    translation~\cite{wirth2001shape}.
    
    \section{Real Data Analysis}
    \label{s:analysis}
    
    We studied the association between tumor shape and prognostic outcomes by applying univariate and multivariate survival analysis techniques to real data. Section~\ref{sb:survival} describes the univariate and multivariate survival analyses, while Section~\ref{sb:performance} concludes the chapter by developing and validating a prognostic model to predict patient survival outcomes.
    
    Two independent cohorts provided the pathology images with the corresponding clinical data. Specifically, the National Lung Screening Trial (NLST) provided $267$ $40\times$ images for $143$ lung ADC patients, while The Cancer Genome Atlas (TCGA) program provided $457$ $40\times$ images for $318$ lung ADC patients.\footnote{The NLST website ({https://biometry.nci.nih.gov/cdas/learn/nlst}) provides additional information on how the images, based on H\&E-stained slides from blocks of tissues, were obtained.} We considered the NLST dataset for initial analyses and training of the prognostic model while the TCGA dataset was used only for validation purposes. Table~\ref{t:characteristics} presents a clinical summary of the patients in each cohort.
    \begin{table}[H]
    	\centering
    	\footnotesize
    	\caption{Patient characteristics of training and validation datasets. Hypothesis testing was performed to test the differences in the two patient cohort; two sample $t$-test (unequal variances) was used for age and chi-square test for remaining variables.}
    	\begin{center}
    		\begin{tabular}[t]{lllr}
    			\toprule
    			& NLST (training) & TCGA (validation) & $p$-value\\
    			\midrule
    			Number of Patients & 143 & 318 & \\
    			Age & 64.01$\pm$5.19 & 64.52$\pm$10.28 & 0.4819\\
    			Gender &  &  & $\bm{0.0140}$\\
    			\hspace{1em}Male & 80 (55.9\%) & 137 (43.1\%) & \\
    			\hspace{1em}Female & 63 (44.1\%) & 181 (56.9\%) & \\
    			Smoking Status &  &  & $\bm{0.0032}$\\
    			\hspace{1em}Yes & 75 (52.4\%) & 214 (67.3\%) & \\
    			\hspace{1em}No & 68 (47.6\%) & 104 (32.7\%) & \\
    			Stage &  &  & $\bm{0.0066}$\\
    			\hspace{1em}I & 95 (66.4\%) & 174 (54.7\%) & \\
    			\hspace{1em}II & 15 (10.5\%) & 78 (24.5\%) & \\
    			\hspace{1em}III & 23 (16.1\%) & 45 (14.2\%) & \\
    			\hspace{1em}IV & 10 (7\%) & 21 (6.6\%) & \\
    			\bottomrule
    			\multicolumn{4}{l}{%
    				\rule{0pt}{1em}Values are either mean $\pm$ standard deviation, or number (percentage).
    			} \\
    		\end{tabular}
    	\end{center}
    	\label{t:characteristics}
    \end{table}
    
    The automated tumor recognition system, based on a convolutional neural network (CNN) and developed in \citet{wang2018comprehensive}, created a reconstructed heat map from each pathology image. Furthermore, each reconstructed heat map was processed using the procedure in Section~\ref{s:segmentation}, resulting in their shape representations and shape-based features. The aforementioned $30$ features were computed for each tumor region. Only the primary tumor regions were kept, that is, the largest tumor region in the largest tissue sample of each slide image.
    
    We note that, for each analysis, the response is overall survival, defined as the period from diagnosis until death or last contact. Additionally, whenever possible, we adjust for clinical variables (age, gender, smoking status, and stage) and account for patients with multiple slide images by clustering patient identifiers. If clustering is unavailable, we summarize the features at the patient-level, using the median. Survival analysis was performed with \texttt{R} software, version 4.02, and \texttt{R} packages \texttt{survival} (version 3.2-7) and \texttt{glmnet} (version 4.0-2) \citep{glmnet-survival,R,survival-package}. The results were considered significant if the two-tailed p-value $\leq 0.05$ and the analyses were based on the methodology in \citet{wang2018comprehensive}.
    
    \subsection{Survival Analysis}
    \label{sb:survival}
    
    We fit a univariate Cox proportional-hazards model (CoxPH) to each scaled feature\footnote{Features were scaled (standardized) by their respective mean and standard deviation.}, adjusting for age, gender, smoking status, and stage as well as clustering by patient identifiers, to study the association between each individual feature and prognostic outcome. The results, using the NLST dataset, are summarized in Table~\ref{t:univariate}. Only eight features were statistically significant (p-value $\leq 0.05$) with patient survival outcome. Out of the eight, seven were shape features, one was a topological feature, while none were boundary features. All of these features were correlated with poor survival outcome with hazard ratios (HRs) $> 1$). Serving as a negative control, the aspect ratio and major axis angle were not correlated with patient prognosis ($p$-value $> 0.05$; Table~\ref{t:univariate}).
    \begin{table}[hbtp!]
    	\caption{Univariate analysis of shape-based features in NLST training dataset. A Cox proportional-hazards (CoxPH) model was fitted to each scaled feature, clustering by patient identifier (Patient ID) and adjusting for clinical variables (see Section~\ref{sb:survival}). }
    	
    	\footnotesize
    	
    	\resizebox{\textwidth}{!}{%
    		\begin{tabular}{lrcccr}
    			\toprule
    			& Coefficient & Hazard ratio (HR) & Standard error (SE) & Robust SE & $p$-value\\
    			\midrule
    			Aspect Ratio & -0.0403 & 0.9605 & 0.1177 & 0.1059 & 0.7031\\
    			Net Area & 0.2547 & 1.2901 & 0.0946 & 0.1321 & 0.0538\\
    			Elongation & -0.2347 & 0.7908 & 0.1613 & 0.1416 & 0.0976\\
    			{Thickness} & {0.2577} & {1.2939} & {0.1017} & {0.1236} & \textbf{0.0370}\\
    			{Number of Holes} & {0.2432} & {1.2753} & {0.0920} & {0.0974} & \textbf{0.0125}\\
    			{Polygon-Based Perimeter} & {0.3404} & {1.4055} & {0.1178} & {0.1380} & \textbf{0.0136}\\
    			{Polygon-Based Area} & {0.2631} & {1.3009} & {0.0962} & {0.1230} & \textbf{0.0324}\\
    			Compactness & 0.0349 & 1.0355 & 0.1263 & 0.1223 & 0.7756\\
    			Roundness & 0.1258 & 1.1340 & 0.1212 & 0.1357 & 0.3539\\
    			Circularity & 0.0857 & 1.0895 & 0.1271 & 0.1531 & 0.5754\\
    			Convexity & 0.0800 & 1.0833 & 0.1244 & 0.1428 & 0.5752\\
    			Density & 0.2305 & 1.2593 & 0.1332 & 0.1528 & 0.1314\\
    			{Major Axis Length} & {0.3738} & {1.4532} & {0.1148} & {0.1363} & {0.0061}\\
    			Major Axis Angle & -0.0744 & 0.9283 & 0.1117 & 0.1147 & 0.5169\\
    			{Minor Axis Length} & {0.2767} & {1.3188} & {0.1132} & {0.1317} & \textbf{0.0356}\\
    			Rectangularity & 0.1342 & 1.1436 & 0.1213 & 0.1440 & 0.3514\\
    			Eccentricity & -0.2324 & 0.7926 & 0.1168 & 0.1207 & 0.0542\\
    			{Fibre Length} & {0.2481} & {1.2816} & {0.1023} & {0.1182} & \textbf{0.0358}\\
    			{Fibre Width} & {0.3336} & {1.3959} & {0.1178} & {0.1375} & \textbf{0.0153}\\
    			Curl & 0.0238 & 1.0240 & 0.1215 & 0.1198 & 0.8428\\
    			Total Absolute Curvature & 0.1121 & 1.1186 & 0.1200 & 0.1628 & 0.4912\\
    			Bending Energy & 0.1099 & 1.1162 & 0.1203 & 0.1623 & 0.4984\\
    			Radial Mean & 0.0862 & 1.0901 & 0.1226 & 0.1343 & 0.5209\\
    			Radial SD & -0.0670 & 0.9352 & 0.1188 & 0.1161 & 0.5642\\
    			Radial-Based Circularity & 0.0720 & 1.0747 & 0.1172 & 0.1195 & 0.5466\\
    			Radial Entropy & -0.0507 & 0.9506 & 0.1174 & 0.1118 & 0.6503\\
    			Area Ratio & -0.0474 & 0.9537 & 0.1203 & 0.1226 & 0.6992\\
    			Zero-crossing Count & 0.1040 & 1.1096 & 0.1127 & 0.1275 & 0.4146\\
    			Average Roughness Index & -0.1156 & 0.8908 & 0.1552 & 0.2609 & 0.6577\\
    			Margin Fluctuation & -0.0669 & 0.9353 & 0.1104 & 0.1203 & 0.5785\\
    			\bottomrule
    			\multicolumn{6}{l}{\rule{0pt}{1em}Bolding signifies features with $p$-value $\leq 0.05$.}\\
    		\end{tabular}
    	}
    	
    	\label{t:univariate}
    \end{table}
    
\begin{figure}[H]
    	\centering
    		\includegraphics[width=0.9\linewidth]{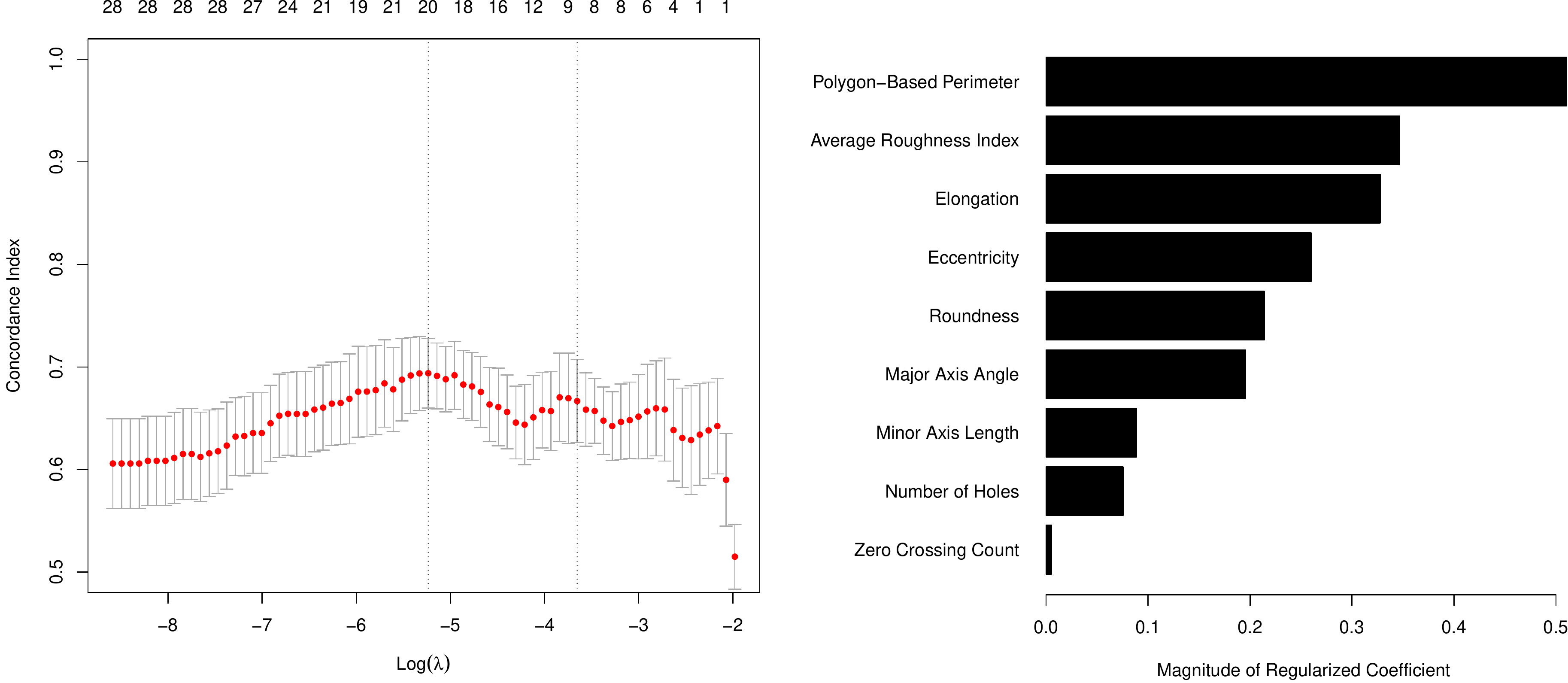}
    	\caption{Results from the regularized Cox proportional-hazards (CoxPH) model. The left figure shows the mean cross-validated errors, based on Harrell's concordance measure, for each penalty coefficient $\lambda$. The right figure shows the importance of each feature kept by the regularized CoxPH model, based on the magnitude of each coefficient.}
    	\label{f:regularization}
    \end{figure}

    All shape-based features, scaled like in the univariate analysis, were selected to build a multivariate Cox proportional-hazards model with a lasso penalty; using Harrell's concordance measure, the penalty coefficient $\lambda$ was determined through $10$-fold cross validation \citep{glmnet-survival,wang2018comprehensive}. We chose the penalty coefficient $\lambda$ such that this value was within $1$ standard error of the value that gave the minimum mean cross-validated error \citep{glmnet-survival}. The mean cross-validated errors for various penalty coefficients $\lambda$ are shown in Figure~\ref{f:regularization} (left).
    
    Similarly, we summarized the standardized coefficients kept by the regularized CoxPH model in Table~\ref{t:multivariate}. Scaling the feature before fitting the model, allowed us to obtain the standardized coefficients. This allowed us to compare the regularized coefficients with the coefficients obtained in the univariate analysis (Table~\ref{t:univariate}). Additionally, we can also show the importance of each feature in Figure~\ref{f:regularization} (right), based on the magnitudes of each regularized coefficient. Clustering of patient identifiers was not used in the regularized Cox proportional-hazards model; instead, we summarized the features at the patient-level. The shape-based prognostic model, based on the predicted risk score of the regularized CoxPH model, was developed, and its predictive performance validated as follows.
    
    \begin{table}[hbtp!]
    	\caption{Multivariate analysis of shape-based features in NLST training dataset. A regularized Cox proportional-hazards (CoxPH) model with a lasso penalty was fitted to each scaled feature. The penalty coefficient $\lambda$ was chosen based on $10$-fold cross validation. The resulting coefficient values can be compared to the univariate coefficients summarized in Table~\ref{t:univariate}.}
    	
    	\footnotesize
    	
    	\begin{center}
    		\begin{tabular}{lrr}
    			\toprule
    			& Coefficient & exp(Coefficient)\\
    			\midrule
    			Aspect Ratio & -- & --\\
    			Net Area & -- & --\\
    			Elongation & -0.3276 & 0.7206\\
    			Thickness & -- & --\\
    			Number of Holes & 0.0754 & 1.0783\\
    			Polygon-Based Perimeter & 0.5106 & 1.6663\\
    			Polygon-Based Area & -- & --\\
    			Compactness & -- & --\\
    			Roundness & 0.2139 & 1.2385\\
    			Circularity & -- & --\\
    			Convexity & -- & --\\
    			Density & -- & --\\
    			Major Axis Length & -- & --\\
    			Major Axis Angle & -0.1956 & 0.8223\\
    			Minor Axis Length & 0.0888 & 1.0929\\
    			Rectangularity & -- & --\\
    			Eccentricity & -0.2600 & 0.7710\\
    			Fibre Length & -- & --\\
    			Fibre Width & -- & --\\
    			Curl & -- & --\\
    			Total Absolute Curvature & -- & --\\
    			Bending Energy & -- & --\\
    			Radial Mean & -- & --\\
    			Radial SD & -- & --\\
    			Radial-Based Circularity & -- & --\\
    			Radial Entropy & -- & --\\
    			Area Ratio & -- & --\\
    			Zero-crossing Count & 0.0054 & 1.0054\\
    			Average Roughness Index & 0.3468 & 1.4145\\
    			Margin Fluctuation & -- & --\\
    			\bottomrule
    		\end{tabular}
    	\end{center}
    	
    	\label{t:multivariate}
    \end{table}
    
    \subsection{Predictive Performance}
    \label{sb:performance}

\begin{figure}[H]
    	\begin{center}
    		\includegraphics[width=0.5\linewidth]{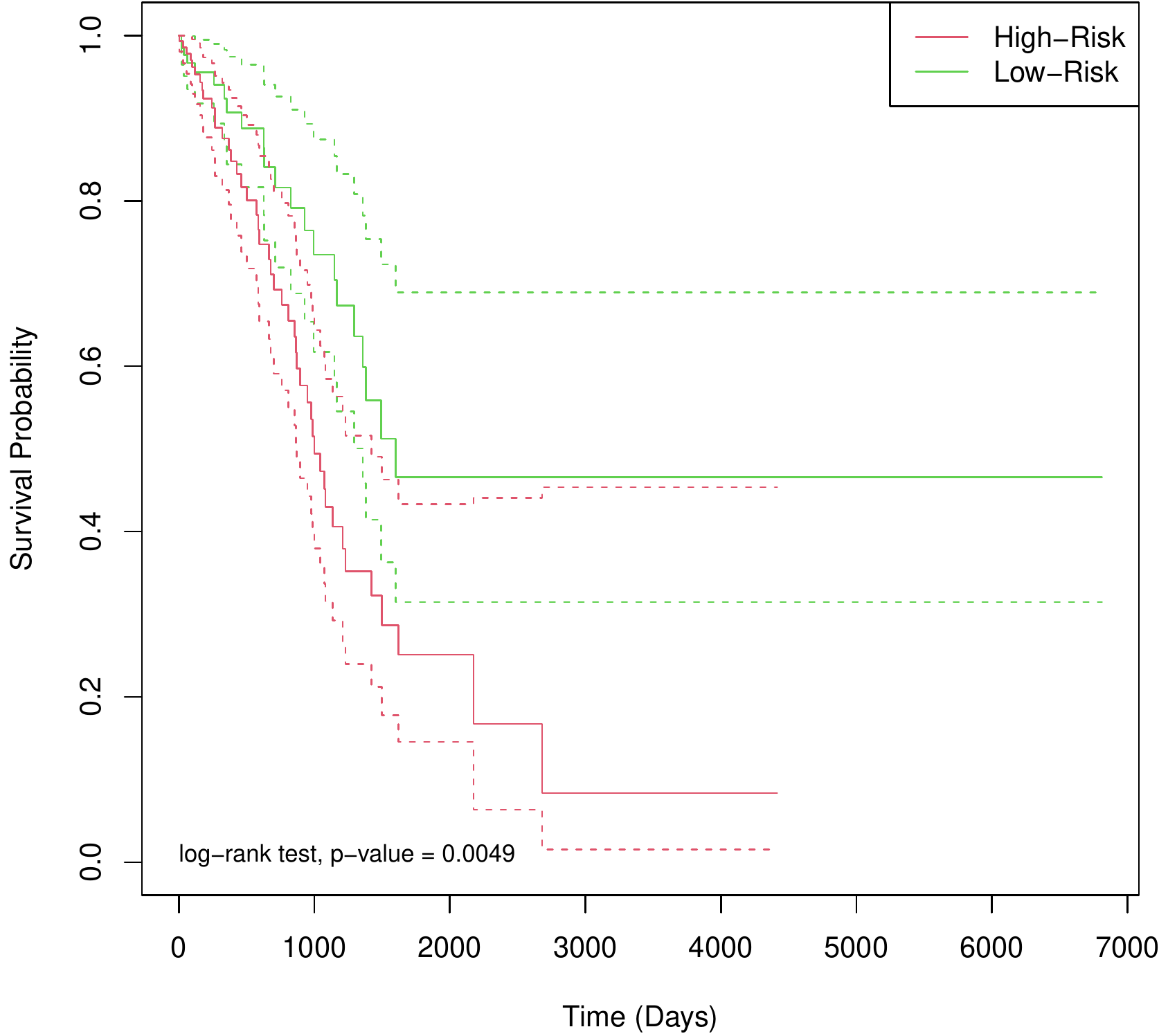}
    	\end{center}
    	\caption{Survival curves, estimated using the Kaplan-Meier method, for both high-risk and low-risk groups. A $\log$-rank test was performed to test the difference between these two curves. The $p$-value of the $\log$-rank test is overlaid and the $95\%$ pointwise confidence intervals are also shown for each risk group.}
    	\label{f:curves}
    \end{figure}

    Since the multivariate model in Section~\ref{sb:survival} was based on the patient-level features, similar processing was applied to the TCGA cohort. Additionally, the features were centered and scaled based on the means and standard deviations of the features in the NLST cohort. These transformations prevent bias and allow us to independently validate the prognostic performance of this model. A shape-based risk score was computed for each patient to assign patients into predicted high-risk and low-risk groups, using the median as a cutoff~\cite{wang2018comprehensive}. The survival curves for the predicted high-risk and low-risk groups were estimated using the Kaplan-Meier method and are shown in Figure~\ref{f:curves}, as well as the results of the $\log$-rank test to compare the survival difference between risk groups. We associate a worse survival outcome with the predicted high-risk group ($\log$-rank test, $p$-value $= 0.0049$, Figure~\ref{f:curves}).
    
    The prognostic performance of the shape-based risk score was validated by a multivariate CoxPH model and the resulting coefficients are summarized in Table~\ref{t:prognosis}. After adjusting for clinical variables, including age, gender, smoking status, and stage, the predicted risk groups independently predicted prognosis (high-risk vs. low-risk, HR $= 2.11$, CI $1.24$--$3.59$, $p$-value $= 0.0059$, Table~\ref{t:prognosis}).
    \begin{table}[H]
        \caption{Multivariate analysis of the predicted risk group. A Cox proportional-hazards (CoxPH) model was fitted to test the predictive performance of the predicted risk group, adjust for clinical variables and based on the TCGA cohort.}
    
        \footnotesize
    
        \begin{center}
            \begin{tabular}{lcr}
                \toprule
                  & Hazard Ratio (HR) with $95\%$ confidence interval (CI) & $p$-value\\
                \midrule
                {High-Risk vs. Low-Risk} & {2.11 (1.24--3.59)} & \textbf{0.0059}\\
                Age & 1.02 (0.99--1.04) & 0.2008\\
                Female vs. Male & 1.66 (0.94--2.96) & 0.0829\\
                Smoker vs. Non-Smoker & 0.81 (0.48--1.36) & 0.4200\\
                {Stage II vs. Stage I} & {2.48 (1.34--4.59)} & \textbf{0.0038}\\
                {Stage III vs. Stage I} & {4.98 (2.59--9.61)} & \textbf{$\boldsymbol{\leq 0.001}$}\\
                {Stage IV vs. Stage I} & {3.48 (1.46--8.28)} & \textbf{0.0047}\\
                \bottomrule
                \multicolumn{3}{l}{%
                    \rule{0pt}{1em} Bolding signifies features with p-value $\leq 0.05$.
                } \\
            \end{tabular}
        \end{center}
    
        \label{t:prognosis}
    \end{table}

    \section{Discussion and Conclusion}
    \label{s:conclusion}
    
    A whole-slide image processing procedure was developed to extract tumor regions and compute shape-based descriptors from them. This procedure facilitated the analyses performed to study the association between tumor shape and patient survival outcome. While this procedure successfully extracted and visualized tumor regions, further inspection is necessary to ensure the extracted objects accurately represent the tumors in the whole-slide image. Specifically, various morphological operations \citep{gonzalez2020digital} could be used in the procedure to improve the main tumor representation.
    
    The univariate analysis discovered eight statistically significant features (at the $5\%$ level of significance); of these eight features, seven attempted to quantify the size and length of the tumor: thickness, polygon-based perimeter and area, major and minor axis length, and fibre length and width (Table~\ref{t:univariate}). The only topological feature proved to be significant: the number of holes, which quantifies the degradations within the tumors. We note that none of these seven features were based on the derived shape representations; specifically, the thickness feature was based on the binary matrix, while the remaining features were based on the polygon chain. Therefore, with the features used, the derived shape representations were ineffective in diversifying the tumor shape. Interestingly, the results align with some of the significant features found in \citet{wang2018comprehensive} such as the area, perimeter, major axis length, minor axis length, and the number of holes.
    
    The multivariate analysis used regularization to avoid overfitting and selected nine features: perimeter, average roughness index, elongation, eccentricity, roundness, major axis angle, minor axis length, number of holes, and zero-crossing count (ordered from largest to smallest importance; Figure~\ref{f:regularization}). Out of the nine features three were significant in the univariate analysis: perimeter, minor axis length, and the number of holes. Interestingly, the major axis angle, a negative control, was kept in the regularized model. Furthermore, the radial-based features, average roughness index, and zero-crossing count, insignificant in the univariate analysis, were also kept by the model. This indicates that further research regarding the derived shape representations and their features should be conducted. The resulting regularized model was used to compute patient risk scores to, subsequently, dichotomize patients into high-risk and low-risk groups. Through an independent cohort, the predicted risk score proved to be a significant prognostic factor, alongside other clinical variables (Table~\ref{t:prognosis}).
    
    Tumor regions obtained from whole-slide images tend to have a complex shape and boundary. These complexities can, potentially, introduce noise into the shape representations. As a result, this noise disrupts the computed features. This issue could be resolved by introducing noise-resistant shape representations and features. We also cannot disregard the $3$-dimensional spatial information that is lost due to two-dimensional imaging procedures. Therefore, we can apply feature extraction to other imaging techniques, such as computed tomography (CT) colonography, magnetic resonance imaging (MRI), and positron emission tomography (PET)/CT colonography, to produce more comprehensive and diverse shape-based features that further quantify the tumor shape, geometry, and topology as well as improving the performance of current risk-based prognostic models \citep{wang2018comprehensive}.
    
    In this study, we developed a whole-slide image processing procedure and shape-based prognostic model to evaluate the relationship between tumor shape, geometry, and topology and patient survival outcome. Our image processing procedure, efficiently, extracted tumor regions, and computed shape-based features from pathology images. This procedure can be adapted to include more shape representations and features; it can also be adapted to other cancer types such as the brain, breast, and kidney cancer. The univariate and multivariate provided new insights into the relationship between tumor shape and patient prognosis. Additionally, our results supported the findings in \citet{wang2018comprehensive}. Our shape-based prognostic model, adapted from \citet{wang2018comprehensive}, predicted patient risk scores. These risk scores served as a prognostic factor, independent of other clinical variables. In all, the proposed processing and model development pipeline provided an objective prognostic method as well as contributing to the shape analysis of tumors.
    
    
    \appendix
    
     \section{Additional Formulas}
    \label{a:formulas}
    
    \subsection{Centroid}
    \label{a:centroid}
    
    The centroid of a shape $\cent = (x_c, y_c)$, also known as the center of
    gravity, is given by
    \begin{align*}
    x_c &= \frac{1}{6A} \sum_{i=1}^n (x_i + x_{i+1}) (x_i y_{i+1} - x_{i+1} y_i) \\
    y_c &= \frac{1}{6A} \sum_{i=1}^n (y_i + y_{i+1}) (x_i y_{i+1} - x_{i+1} y_i)
    \end{align*}
    where
    \[
    A = \frac{1}{2} \sum_{i=1}^n (x_i y_{i+1} - x_{i+1} y_i)
    \]
    is the signed area of the shape, obtained using Gauss's area formula.
    For a region, represented in a two-dimensional discrete plane, the
    centroid $\bar{c} = (\bar{l}, \bar{w})$ is given by
    \[
    \bar{l} = \frac{1}{N} \sum_{l} \I_R(l, w) \cdot l \quad \text{and} \quad
    \bar{w} = \frac{1}{N} \sum_{w} \I_R(l, w) \cdot w
    \]
    where $N = \sum_{l,w} \I_R(l, w)$ is the number of points that make up
    the region.\footnote{
    	We can see that the coordinates of the centroid are the means
    	of the $xy$-coordinates that make up the region.
    }
    
    
    

    \newpage
    \bibliographystyle{imsart-nameyear}
    \bibliography{bibliography}
\end{document}